\documentclass[prl,
superscriptaddress,longbibliography,11pt
]{revtex4-1}

\usepackage{lineno}
\usepackage{amsmath,amssymb,amsthm,amsfonts,mathrsfs}
\usepackage{bm}
\usepackage{graphicx}
\usepackage{color}
\usepackage{tikz,pgf}
\usetikzlibrary{shadings}

\usepackage{dsfont}
\usepackage{fontenc}
\usepackage{amsfonts}
\usepackage{booktabs}
\usepackage[colorlinks=true, letterpaper=true, pdfstartview=FitV, linkcolor=blue, citecolor=blue, urlcolor=blue]{hyperref}

\def \k {\mathbf{k}}
\def \A {\mathcal{A}}
\def \B {\mathcal{B}}

\begin{document}

\title{Circumventing the no-go theorem: A single Weyl point surrounded by nodal walls}

\author{Zhi-Ming Yu}
\thanks{Z.-M. Yu and W. Wu contributed equally to this work}
\address{Research Laboratory for Quantum Materials, Singapore University of Technology and Design, Singapore 487372, Singapore}

\author{Weikang Wu}
\thanks{Z.-M. Yu and W. Wu contributed equally to this work}
\address{Research Laboratory for Quantum Materials, Singapore University of Technology and Design, Singapore 487372, Singapore}

\author{Y. X. Zhao}
\email{zhaoyx@nju.edu.cn}
\address{National Laboratory of Solid State Microstructures and Department of Physics, Nanjing University, Nanjing 210093, China}
\address{Collaborative Innovation Center of Advanced Microstructures, Nanjing University, Nanjing 210093, China}

\author{Shengyuan A. Yang}
\email{shengyuan\_yang@sutd.edu.sg}
\address{Research Laboratory for Quantum Materials, Singapore University of Technology and Design, Singapore 487372, Singapore}

\begin{abstract}
Despite of a rapidly expanding inventory of possible crystalline Weyl semimetals, all of them are constrained by the Nielsen-Ninomiya no-go theorem, namely, that left- and right-handed Weyl points appear in pairs. With time-reversal ($\mathcal{T}$ ) symmetry, an even stronger version holds for the semimetals, \emph{i.e.}, all eight time-reversal-invariant points in the Brillouin zone (BZ) simultaneously host Weyl points or not. However, all the well-known conclusions from the no-go theorem are implicitly within the current framework of topological semimetals, and in this Letter we shall go beyond it by exploring composites of topological metal and semimetal phases.  Guided by crystal symmetry and $\mathcal{T}$ symmetry, we propose a new topological phase of $\mathcal{T}$-invariant crystalline metal, where a single Weyl point resides at the center of the BZ, surrounded by nodal walls spreading over the entire BZ boundary. In other words, a single Weyl point is realized with the no-go theorem being circumvented. Meanwhile, the Fermi arc surface states, considered as a hallmark of Weyl semimetals, do not appear for this new composite topological phase. We show that this phase can be realized for space group 19 and 92, with and without spin-orbit coupling, respectively.	
\end{abstract}

\maketitle

Weyl fermions are named after Hermann Weyl \cite{Weyl1929}, who first proposed them as more elementary fermions to preserve the Lorentz symmetry than Dirac fermions, since a pair of Weyl fermions with opposite chirality can be hybridized to compose a Dirac fermion. Although the long endeavor in high energy physics to search elementary particles as Weyl fermions has no result so far,
Weyl fermions have been found in condensed matter physics, as emergent quasi-particles, first in the A phase of $^3$He superfluid \cite{Volovik2003}, and recently in various crystalline solids known as Weyl semimetals \cite{wan2011topological,murakami2007phase}, which become a hot topic in current research \cite{Armitage_RMP}. It is now clear that their ubiquitous emergence is due to their topological charge, namely the unit Chern number $\pm 1$ associated with Weyl points, where they are excited in momentum space. On the other hand, the topological charge leads to the Nielsen-Ninomiya no-go theorem \cite{NIELSEN1981I,NIELSEN1981II}, which asserts that left- and right-handed Weyl points always appear in pairs and hence excludes the possibility of only a single Weyl point existing in momentum space of a Weyl semimetal.

\begin{figure}[h]
	\includegraphics[scale=0.4]{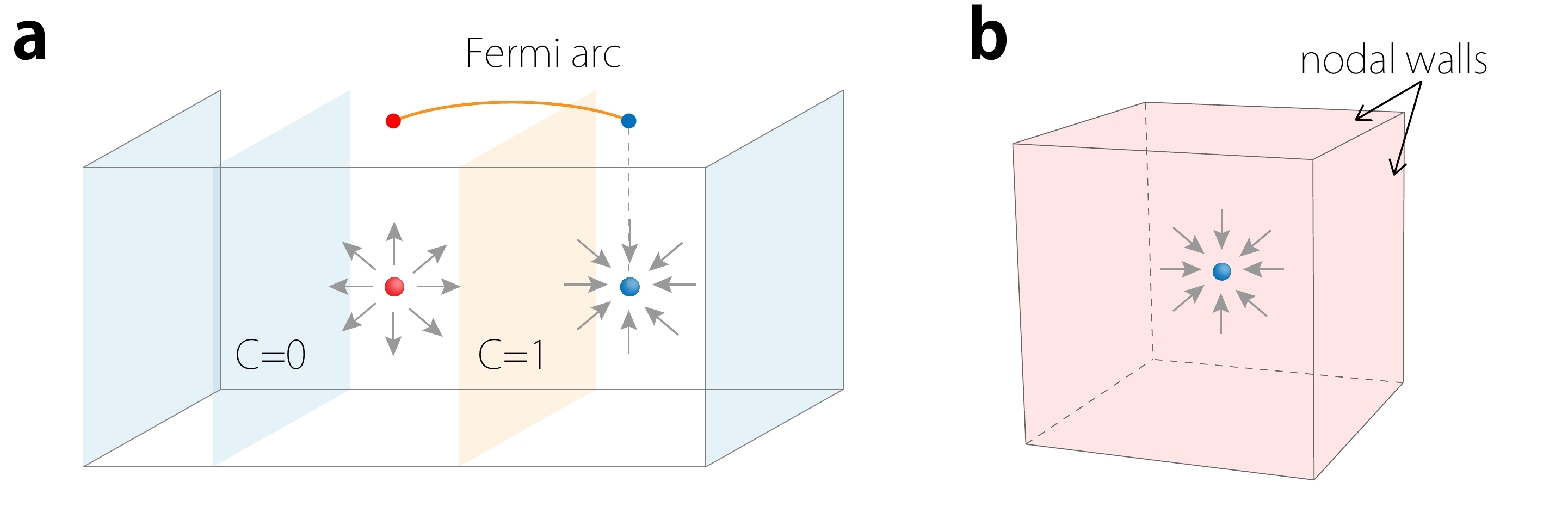}
	\caption{\textbf{Illustration of the strategy to circumvent the no-go theorem.} \textbf{a}, Conventional Weyl semimetal respecting no-go theorem and exhibiting Fermi arc surface states. When a gapped 2D slice in the BZ (the shaded region) moves across a Weyl point, its Chern number changes by a unit. The periodicity of the BZ requires at least another Weyl point with the opposite chirality. The chiral edge modes for the nontrivial 2D slices trace out the Fermi arc on the surface, connecting the surface projection of the two Weyl points.  \textbf{b}, Schematic of the new topological metal phase proposed here. A single Weyl point sitting inside the BZ is surrounded by nodal walls covering the entire BZ boundary. For such case, one cannot find any gapped 2D slices of the BZ, so the argument in \textbf{a} no longer applies.  \label{fig_illust}}
\end{figure}

A simple argument for the no-go theorem goes as following. In a band theory with well separated valence and conduction bands except at isolated Weyl points, a ``magnetic field'' in the momentum space known as Berry curvature can be defined as $\B(\k)=\nabla\times\A(\k)$, where the Berry gauge potential $\A(\k)=\sum_a i\langle a,\k|\nabla_\k|a,\k\rangle$ with $a$ labeling the valence bands \cite{Niu2017book}. The Weyl points are just monopoles for this $\B$ field, and according to Gauss law, the charge of a Weyl point is just the integer-quantized total flux it emits in units of $2\pi$, which coincides with its chirality.
Now consider a 2D slice in the 3D Brillouin zone (BZ). The Chern number of this slice is changed by a unit when it moves across a Weyl point, say, from 0 to 1 as illustrated in Fig. \ref{fig_illust}(a). Because the BZ is periodic, the slice must cross another Weyl point with opposite chirality before it can return to its original position with Chern number 0. This argument also demonstrates the existence of topological surface states: The slices with a nontrivial Chern number each has a chiral edge mode, therefore, a Fermi arc composed of these edge modes must exist on the system surface and connect the pair of Weyl points \cite{wan2011topological}. The Fermi arc surface states have been regarded as a hallmark of Weyl semimetals [see Fig. \ref{fig_illust}(a)].

In lattice gauge theory of high energy physics, it has been tried to circumvent the no-go theorem to study Weyl fermions with a single chirality
by resorting to higher dimensions, e.g., by considering the $(3+1)$D boundary of a $(4+1)$D topological insulator \cite{Jackiw1976,Kaplan1992,Kaplan09,luscher2001chiral}.
Here, we reveal a new possibility in condensed matter. It is enabled by noticing a loophole in the above argument: Chern number is only well defined when the 2D slice has a gapped spectrum, then, what if \emph{all} possible BZ slices are \emph{not} gapped?
This is possible because condensed matter systems are not constrained by the Lorentz symmetry, and therefore have more possibilities of gapless manifolds compared to high-energy models. Particularly, gapless points can form closed manifolds in the BZ, such as nodal lines and walls (also named as nodal surfaces) \cite{YangPRL2014,WengPRB2015,ZhongNs2016,LiangPRB2016,Nodal-Surface}.
In this article, we propose a new $\mathcal{T}$-invariant crystalline topological metal phase that circumvents the no-go theorem. It has a single Weyl point residing at the center of the BZ, completely enclosed by nodal walls spreading over the entire BZ boundary, as illustrated in Fig. \ref{fig_illust}(b). Because of the nodal walls, the previous argument no longer applies. Thus, the existence of a single Weyl point and the absence of Fermi arc surface states can become topologically allowed.

To realize this, first of all, the nodal walls must be stable. Notably, nodal walls recently have been attracting attentions in the community~\cite{Nodal-Surface}, where topologically and crystal-symmetry-protected nodal surfaces have been systematically studied. Particularly, the combination of time-reversal and a two-fold screw-rotation symmetry can ensure two-band crossings spreading over the entire face of the BZ boundary perpendicular to the rotation axis. For example, consider the screw rotation $S_{2z}=\{C_{2z}|00\frac{1}{2}\}$, which is rotating $\pi$ around the $z$-axis followed by a half lattice translation along $z$. The composite symmetry $\mathcal{T}S_{2z}$ gives rise to a nodal surface at the BZ boundary $k_z=\pi$, because $({\cal T}S_{2z})^{2} = -1$ on the boundary leading to the Kramers degeneracy. Note that $({\cal T}S_{2z})^{2} = -1$ and therefore the degeneracy hold both with and without spin-orbit coupling (SOC), but for the case with SOC, inversion symmetry has to be broken \cite{Nodal-Surface}.

On the other hand, to realize a single Weyl point in the interior of the BZ, one easy approach is to utilize the Kramers degeneracy due to $\mathcal{T}$ at the time-reversal-invariant points of the BZ. There are eight of them: one at the BZ center $\Gamma$, while the other seven at the BZ boundary. Furthermore, there is actually a stronger no-go theorem for $\mathcal{T}$-invariant Weyl semimetals~\cite{Strong-No-Go}, which asserts that all of the eight Kramers-degenerate points must be simultaneously Weyl points or not. Thus, in order to have a single Weyl point, all three faces of the BZ boundary have to host nodal walls, such that all Weyl points can be absorbed into the nodal walls except the one at the center. Otherwise, if there are only two (one) faces host nodal walls, two (four) Weyl points will simultaneously appear at the two (four) time-reversal-invariant points not covered by nodal walls.

Guided by the above considerations, we look for a $\mathcal{T}$-invariant system with space group (SG) containing twofold screw-rotation symmetries for all three directions.
Let's focus on spinful systems with SOC considered. This further requires that the SG does not contain inversion symmetry. One simplest space group that fulfills these requirements is SG 19. It should be noted that though $\mathcal{T}$ symmetry guarantees the Kramers degeneracy at the BZ center, whether it can indeed lead to a Weyl point will depend on the specific case.

Now, we proceed to explicitly demonstrate the new phase with a single Weyl point by constructing a minimal lattice model of SG 19.
This SG requires the model to contain at least eight bands.
The model settles on a tetragonal lattice, as shown in Fig. \ref{SG19_withSOC}(a). Each unit cell contains four sites denoted by the blue balls, and each site has an $s$-like orbital with two spin states. The gray-colored balls represent sites that do not directly enter the lattice model, but they affect the hopping amplitudes between the active (blue-colored) sites in a way that follows the SG symmetry.
Then, we construct the following lattice Hamiltonian allowed by the symmetry:
\begin{eqnarray}\label{Model-SG19}
{{\cal H}} & = & A_0+A_{1}\cos\frac{k_{y}}{2}\left(\cos\frac{k_{z}}{2}\Gamma_{10}+\sin\frac{k_{z}}{2}\Gamma_{20}\right)+A_{2}\cos\frac{k_{x}}{2}\left(2\cos\frac{k_{z}}{2}\Gamma_{01}+\cos\frac{k_{y}}{2}\Gamma_{11}+\sin\frac{k_{y}}{2}\Gamma_{12}\right)\nonumber \\
 &  & -A_{2}\cos\frac{k_{x}}{2}\left(\cos\frac{k_{y}}{2}\Gamma_{12}+\sin\frac{k_{y}}{2}\Gamma_{11}\right)s_{y}+A_{3}\sin\frac{k_{z}}{2}\left(\sin\frac{k_{x}}{2}\Gamma_{32}+\cos\frac{k_{x}}{2}\Gamma_{01}\right)s_{z}\nonumber \\
 &  & +A_{4}\cos\frac{k_{z}}{2}\left(\cos\frac{k_{x}}{2}\Gamma_{32}+\sin\frac{k_{x}}{2}\Gamma_{01}\right)s_{x}+\sin\frac{k_{x}}{2}\left(A_{5}\sin\frac{k_{y}}{2}\Gamma_{12}+A_{6}\cos\frac{k_{y}}{2}\Gamma_{11}\right)s_{x},
\end{eqnarray}
where the Pauli matrices $s_i$'s  act on the spin space, the $4\times 4$ gamma matrices $\Gamma_{ij}=\sigma_{i}\otimes\sigma_{j}$  act on the four sites [$\sigma_i$'s ($i=1,2,3$) are the Pauli matrices and $\sigma_0$ is the $2\times 2$ identity matrix], the wave vectors are measured in units of corresponding inverse lattice constants, and the $A_i$'s are real parameters. The band structure of this lattice model is plotted in Fig.~\ref{SG19_withSOC}(c).

Here we focus on the lowest two bands (e.g., assuming the electron filling is one electron per unit cell). Clearly, they form a linear type crossing at the $\Gamma$  point, which reflects the Kramers degeneracy of $\mathcal{T}$ symmetry. A detailed calculation shows that the Chern number of the valence band (i.e., the lowest band here) on a sphere surrounding the $\Gamma$ point is ${\cal C}=-1$, indicating that the two-band crossing point is indeed a Weyl point with monopole charge (chirality) $-1$. Meanwhile, as required by the three screw axis (combined with $\mathcal{T}$), all of the three BZ faces are observed to be nodal walls, where the two bands are degenerate. Furthermore, careful scanning over the whole BZ shows that the band structure is fully gapped spreading away from the Weyl point at the BZ center until reaching the nodal walls at the BZ boundary.

To further confirm the topological configuration, we plot the distribution of the Berry curvature field $\mathcal{B}(\mathbf{k})$ in the BZ for the valence band in Fig. \ref{SG19_withSOC}(d). It shows that the ``magnetic field'' is weakly emitted from the nodal walls on the BZ boundary, then flows smoothly towards the center of the BZ, gradually becoming stronger and stronger, and eventually converges onto the monopole singularity, namely the Weyl point, at the center of the BZ. This picture vividly mimics the field configuration for a charged particle sitting inside a conducing box.

We now discuss the bulk-boundary correspondence of the topological metal phase. Given the aforementioned exotic topological configuration, it is expected to be in sharp contrast with the well-accepted bulk-boundary correspondence of Weyl semimetals, where every Weyl point is connected with surface Fermi arcs. The presence of nodal walls precludes any gapped BZ slice for the topological argument of surface Fermi surfaces. Hence, it is possible to have no topological Fermi arc surface states despite the existence of Weyl point. Indeed, our numerical calculation confirms this point.
For example, the projected spectra on the both $(010)$ and $(001)$ surfaces in Fig.~\ref{fig_SG19_SS} find no Fermi arc surface states.

As we have mentioned, the SGs  also can host nodal walls in the absence of SOC.
Although SOC exists in all realistic material, ignoring SOC can be a good approximation for materials with light elements.
Therefore, it is also interesting to explore the similar topological metal phase in systems without SOC. Opposite to the previous discussion with SOC, since now $\mathcal{T}^2=1$, there is no Kramers degeneracy at $\Gamma$.  Hence, to pin a Weyl point at the BZ center, additional crystal symmetry is needed. Here we choose a model with SG 92 as an example. The model details are relegated to the Supplementary Information. It again has four sites in a tetragonal unit cell. The nodal walls on the BZ boundary are still protected by $\mathcal{T}S_{2i}$ ($i=x,y,z$), respectively, but the twofold degeneracy for the Weyl point appears as a two-dimensional irreducible representation of $\mathcal{T}S_{4z}$, with $S_{4z}=\{C_{4z}|\frac{1}{2}\frac{1}{2}\frac{1}{4}\}$ a fourfold screw rotation along $z$.

\begin{figure}[h]
	\includegraphics[scale=0.25]{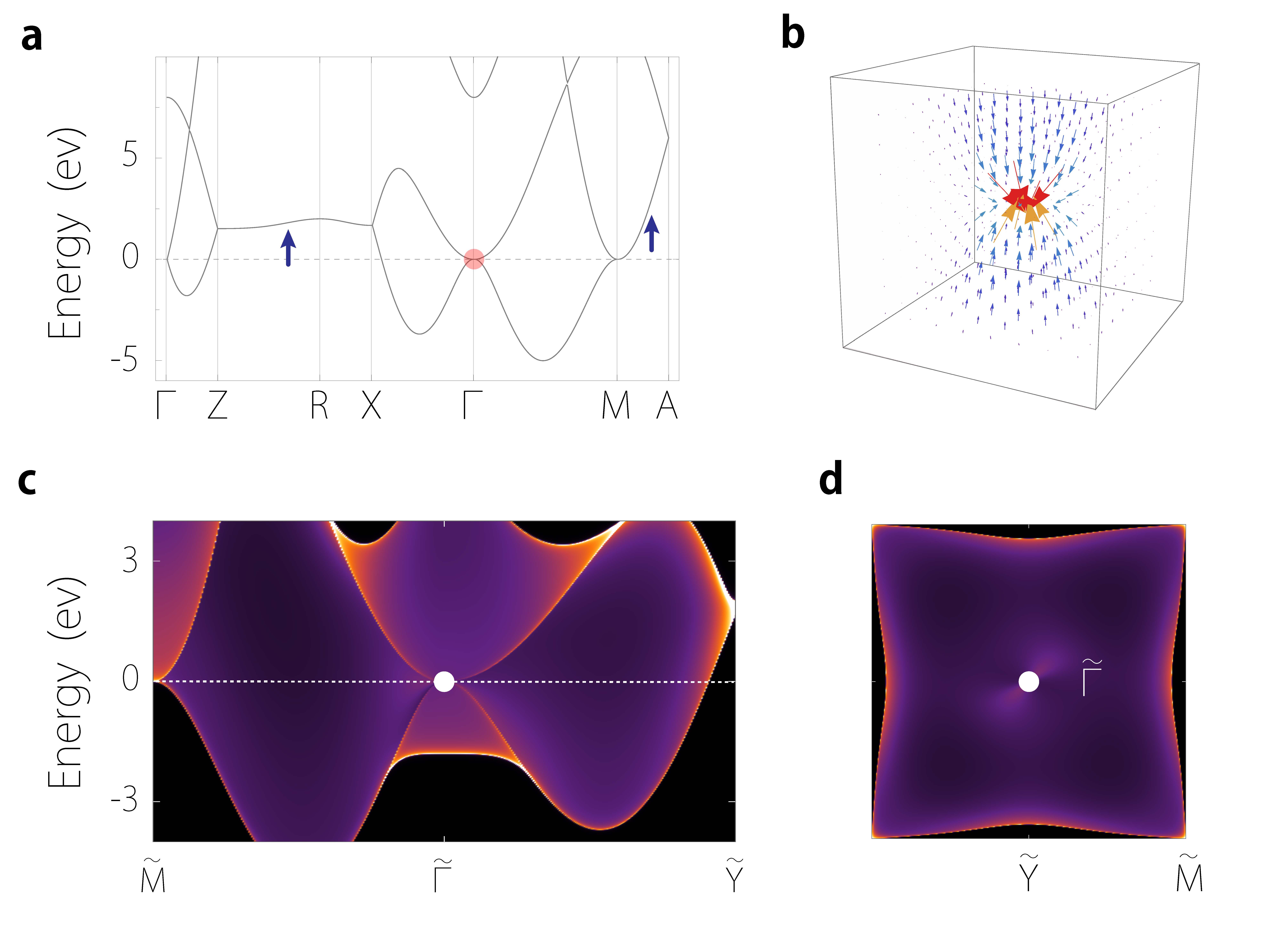}
	\caption{\textbf{A lattice model of SG 92 without SOC.} \textbf{a}, Electronic structures for the model along high symmetry lines. Here we focus on the manifolds of the lowest two bands. The red dot marks the double Weyl point at $\Gamma$, which carries Chern number $-2$. The arrows indicate the degeneracy on the nodal walls. \textbf{b}, Distribution of Berry curvature field in the BZ. \textbf{c}, Projected spectrum on the (001) surface. The projection of the bulk Weyl point is marked by the white dot. \textbf{d} shows the constant energy slice taken at the Weyl point energy in \textbf{c}.  \label{SG92_withoutSOC}}
\end{figure}

The band structure for this model is plotted in Fig.~\ref{SG92_withoutSOC}.
The presence of the fourfold screw rotation only for the $z$-axis reflects the anisotropy of the model, and consistently the dispersion of the Weyl point is quadratic over the $k_x$-$k_y$ plane and linear along $k_z$. Accordingly, the monopole charge is found to be two, namely, that two quanta of flux is ended onto the Weyl point (also known as double Weyl point \cite{XuPRL2011,FangPRL2012}), different from the previous case. The surface states are numerically worked out as well without any Fermi arc states being found [see Fig.~\ref{SG92_withoutSOC}(c,d)].

In this work, we have demonstrated a mechanism to circumvent the famous Nielsen-Ninomiya no-go theorem by discovering a new topological metal phase with a single Weyl point surrounded by nodal walls. We show that this phase may have no Fermi arc surface states, in contrary to the bulk-boundary correspondence for conventional Weyl semimetals. These results are of fundamental significance. The finding opens a new route to study Weyl fermions of a single chirality in condensed matter systems.

We remark that it is also possible to have Weyl points coexisting with one or two nodal walls. As we have discussed, for such cases, there must be more than one Weyl points in the BZ. These phases also exhibit interesting phenomena as we briefly discuss in the following. Consider the case with two nodal walls.
According to the general theory~\cite{Strong-No-Go}, there must be two Weyl points, e.g., at $\Gamma$ and $Z$ in a tetragonal BZ, which can have the same or opposite chirality. If they have the same chirality, the nodal walls must absorb (emit) two flux quanta they emit (absorb), and therefore are topologically charged. Hence, on the (001) surface, two Fermi arcs emerge in the surface BZ, where each connects the coincident projection of two Weyl points to the projection of bulk nodal walls at the surface BZ boundary. On the other hand, if two Weyl points have opposite chirality, they are topologically neutral as a whole, and therefore the nodal walls, although stably exist, have no net topological charge. Accordingly, no Fermi arc is protected on the (001) surface. Thus, in a sense, the (topological) charge at the walls is ``induced" by the charge particles (Weyl points) inside, just like the situation in electromagnetism. The similar discussion can be extended to cases with one nodal wall and to systems without SOC.

\section*{}

\textbf{Methods.}

\textbf{Derivation of minimal lattice model of SG 19.}
We adopt the method in Ref.~\cite{Wieder2016} to construct the minimal lattice (tight-binding) model of SG 19. We first write down the matrix representations for the  generators of SG 19 at $\Gamma$ (including their operations on the wave-vector $\bm{k}$) \cite{Bradley1972}, and then obtain  the symmetry-allowed hopping terms. For SG 19, the presence of three orthogonal twofold screw-rotational symmetries requires the minimal lattice models to have at least eight bands~\cite{WatanabePRL2016}.

\begin{figure}[h]
	\includegraphics[scale=0.25]{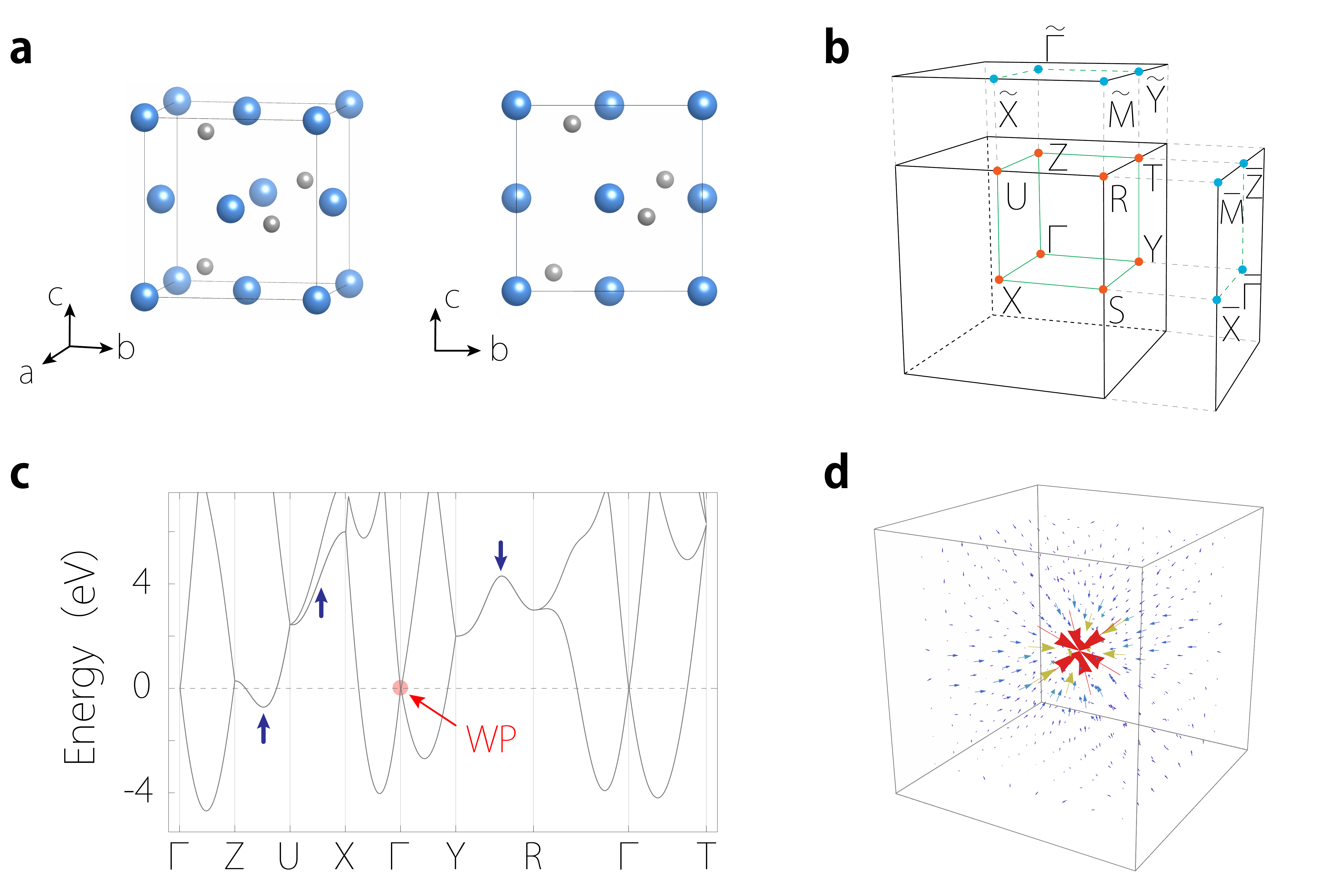}
	\caption{\textbf{A lattice model of SG 19 with SOC.} \textbf{a}, Perspective view and top view of the primitive unit cell of the lattice. The blue balls are the sites with active orbitals, and the gray balls are added to ensure the space group symmetry. \textbf{b}, Brillouin zone of SG 19 and the surface BZs for the (001) and (010) surfaces. \textbf{c}, Electronic band structure of the lattice model. Here we focus on the manifolds of the lowest two bands. The red dot marks the Weyl point at $\Gamma$, which carries Chern number $-1$. The blue arrows indicate the degeneracy on the nodal walls. \textbf{d}, Distribution of Berry curvature field in the BZ. \label{SG19_withSOC}}
\end{figure}

\begin{figure}[h]
	\includegraphics[scale=0.35]{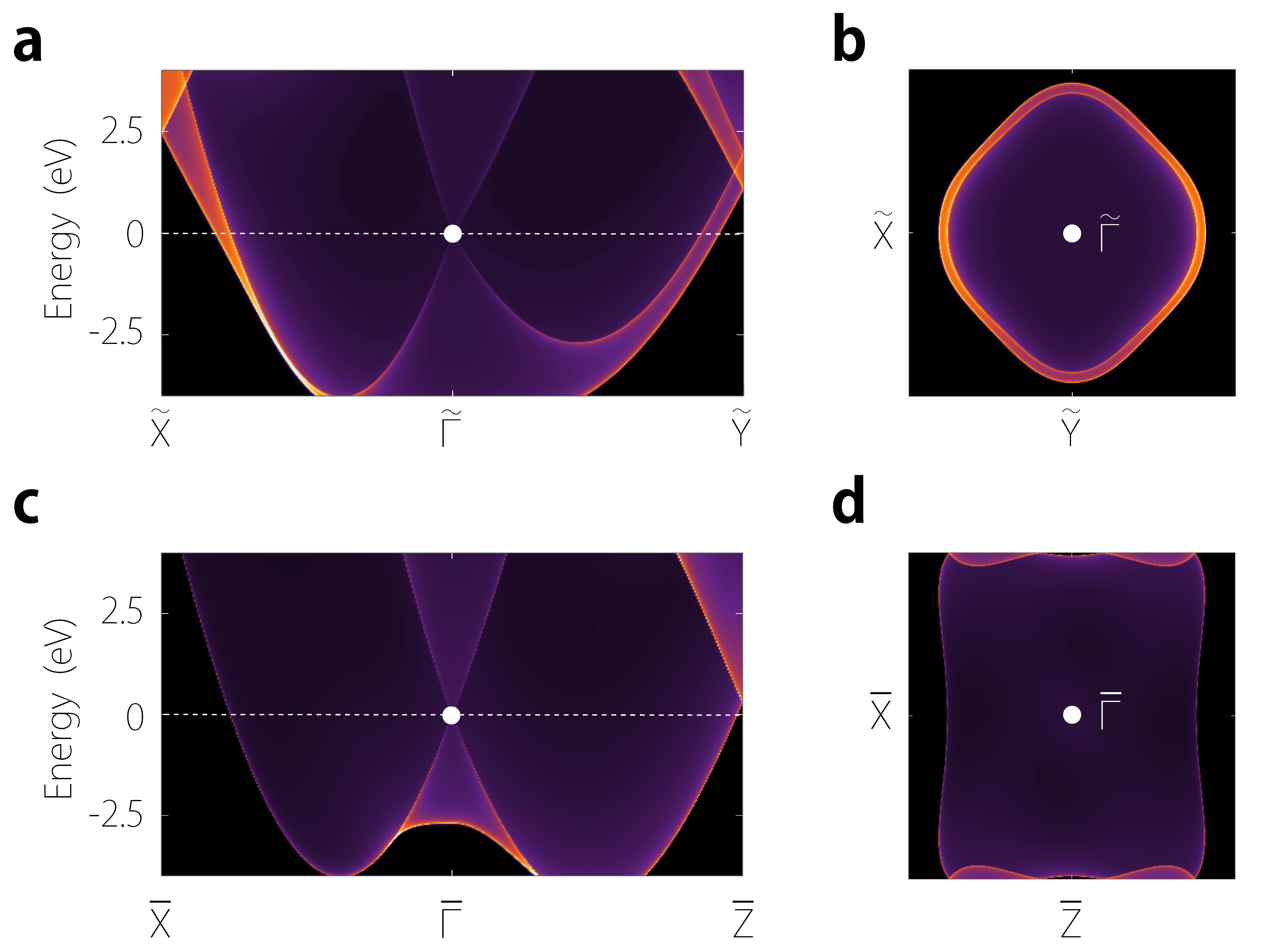}
	\caption{\textbf{Surface spectrum for the lattice model of SG 19 with SOC.} \textbf{a}, Projected spectrum on the (001) surface. The projection of the bulk Weyl point is marked by the white dot. \textbf{b} shows the constant energy slice at the Weyl point energy. \textbf{c} and \textbf{d} show the corresponding results for the (010) surface. No Fermi arc surface state is observed in the spectra.  \label{fig_SG19_SS}}
\end{figure}

The lattice Hamiltonian for SG 19 with SOC is constrained by the following symmetry generators: $S_{2z}=\left\{ C_{2z}\big|\frac{1}{2}0\frac{1}{2}\right\} $,
$S_{2y}=\left\{ C_{2y}\big|0\frac{1}{2}\frac{1}{2}\right\} $ and ${\cal T}$. Here, we have eight basis states per unit cell $\left\{ A_{1},\ A_{2},\ A_{3},\ A_{4}\right\} \otimes\left\{ |\uparrow\rangle,\ |\downarrow\rangle\right\} $, where $A_{i}$ ($i=1,2,3,4$) denote the four sites at positions $ (0,0,0),\ (\frac{1}{2},0,\frac{1}{2}),\ (0,\frac{1}{2},\frac{1}{2}),\ (\frac{1}{2},\frac{1}{2},0) $ {[}see Fig. \ref{SG19_withSOC}(a){]}. With this basis, the  matrix representations  for the   symmetry operators can be written as
\begin{eqnarray}
S_{2z} & = & i\Gamma_{01}s_{z}\otimes\left(k_{x}\rightarrow-k_{x},k_{y}\rightarrow-k_{y}\right),\\
S_{2y} & = & i\Gamma_{10}s_{y}\otimes\left(k_{x}\rightarrow-k_{x},k_{z}\rightarrow-k_{z}\right),\\
{\cal T} & = & is_{y}{\cal K}\otimes\left(\bm{k}\rightarrow-\bm{k}\right),
\end{eqnarray}
with ${\cal K}$ the complex conjugation operator.
The constructed lattice model is presented in main text as Eq.~(\ref{Model-SG19}). One can check that all the terms in Hamiltonian (\ref{Model-SG19}) is invariant under the above symmetry operations.  The  parameters used for the plots in Fig.~\ref{SG19_withSOC} and Fig.~\ref{fig_SG19_SS}  are $A_{0}=21$ eV, $A_{1}=-2$ eV, $A_{2}=-6$ eV, $A_{3}=-12$ eV, $A_{4}=1$ eV, $A_{5}=-6$ eV and $A_{6}=-14$ eV.


\bibliography{Topo-Wall_refs}

\newpage

\end{document}